\documentclass[aps,preprint,preprintnumbers,amsmath,amssymb,nofootinbib]{revtex4}
\usepackage{amssymb}
\usepackage{lscape}
\usepackage{amsmath}
\usepackage{rotating}
\usepackage{bm}
\usepackage{graphicx}
\usepackage{epsfig}
\begin{document}
\title {Gravito-magnetic monopoles in traversable wormholes from WIMT}
\author{$^{2}$
Jes\'us Mart\'{\i}n Romero\footnote{E-mail address:
jesusromero@conicet.gov.ar}, $^{1,2}$ Mauricio Bellini
\footnote{E-mail address: mbellini@mdp.edu.ar} }
\address{$^1$ Departamento de F\'isica, Facultad de Ciencias Exactas y
Naturales, Universidad Nacional de Mar del Plata, Funes 3350, C.P.
7600, Mar del Plata, Argentina.\\
$^2$ Instituto de Investigaciones F\'{\i}sicas de Mar del Plata (IFIMAR), \\
Consejo Nacional de Investigaciones Cient\'ificas y T\'ecnicas
(CONICET), Argentina.}

\begin{abstract}
Using Weitzenb\"ock Induced Matter Theory (WIMT), we study Schwarzschild wormholes performing different foliations on an extended (non-vaccuum) 5D manifold. We explore the geodesic equations for observers which are in the interior of a traversable wormhole and how these observers can detect gravito-magnetic monopoles which are dual to gravito-electric sources observed in the outer zone of some Schwarzschild Black-Hole (BH). The densities of these monopoles are calculated and quantized in the Dirac sense. This kind of duality on the extended Einstein-Maxwell equations, relates electric and magnetic charges on causally disconnected space regions.
\end{abstract}
 \maketitle

\section{Introduction}

Lorentzian wormholes known as Schwarzschild wormholes or Einstein-Rosen bridges are connections between areas of space that can be modeled as vacuum solutions of the Einstein field equations, that are intrinsic parts of the maximally extended version of the Schwarzschild metric describing an eternal black hole with no charge and no rotation. The Einstein-Rosen bridge was discovered by L. Flamm\cite{flamm} in 1916, a few months after Schwarzschild published his solution and rediscovered by Einstein and Rosen in 1935\cite{ER}. The mathematician H. Weyl proposed the wormhole theory in 1921\cite{weyl}, in the framework of the mass analysis of the electromagnetic energy. In 1962 Wheeler and Fuller shown that this kind of wormhole is unstable if it connects two parts of the same universe\cite{WF}. In the pure Gauss-Bonnet gravity, which is an extension of General Relativity involving extra spatial dimensions, sometimes studied in the context of brane cosmology, wormholes can exist even with no matter\cite{GB}. Stable static solutions of thin-shell wormholes with charge, have been obtained recently in $F(R)$ gravity\cite{EA}. Wormholes were also considered as geometric models of elementary particles, handles of space trapping inside an electric flux, say, which description may indeed be valid at the Planck scale\cite{4}. Wormholes can also describe initial data for the Einstein equation\cite{5,6}, whose time evolution corresponds to the black hole collisions of the type observed in the recent GW150914 event\cite{7}.

On the other hand, the central idea of Induced Matter Theory (IMT)\cite{STM} is that ordinary matter and physical
fields, which are present in our 4D space-time can be geometrically induced by a foliation over a 5D space-time which is at least
Ricci-flat with respect to the Riemannian connections. The theory is formulated taking into account a non-compact extra dimension, but
is compatible with a compact extra dimension too. The 5D physical vacuum is defined by the Ricci null tensor, which implies a zero 5D
Einstein tensor. Such condition must be viewed as a particular case of the hypotheses of the Campbell-Magaard embedding theorem of a 5D Einstein manifold $^{(5D)}Rab = \lambda\,^{(5D)} gab$, with $\lambda = 0$. An extension of this formalism is WIMT\cite{wimt}, which is based in the Weitzenb\"ock's geometry. Some inputs of the Weitzenb\"ock geometry were included in the Appendix (\ref{ap1}). This makes possible the use of IMT-like formalism for any 5D space-time, even if this is not flat in the Riemannian sense.

In this work we study the dual sources from Gravitoelectrodynamics using WIMT inside and outside an effective 4D traversable Schwarzschild wormhole using different foliations on a
5D extended Schwarzschild Black-Hole (BH). This issue is very important because it is possible to show that a gravito-magnetic monopole which is in the interior of an effective 4D wormhole is a dual source to a gravito-electric source, which is outside the horizon of the effective 4D Schwarzschild BH. The paper is organized as follows. In Sect. II we revisit the formalism
of Gravitoelectrodynamics from WIMT. In Sects. III we introduce the traversable wormhole from an extended 5D Schwarzschild metric, and we study the dynamics of the observers which adopt
some specific foliations in can leave in the interior of a traversable wormhole and in the exterior of a Schwarzschild BH.

\section{Gravitoelectrodynamics from WIMT}

The Riemann-Weitzenb\"ock curvature tensor has the same form than the Riemannian curvature tensor, but expressed in terms of Weitzenb\"ock connections, denoted by $^{(W)}\Gamma^a_{dc}$. In a coordinate (holonomic) basis the Riemann-Weitzenb\"ock curvature takes the form
\begin{equation}
^{(W)}R^a_{bcd} = \,^{(W)}\Gamma^a_{dc\, ,\, b} -\,^{(W)}
\Gamma^a_{db\, ,\, c}+\,^{(W)}\Gamma^n_{dc}\,^{(W)}\Gamma^a_{ nb}
-\,^{ (W)}\Gamma^n_{ db}\,^{(W)}\Gamma^a_{nc},
\end{equation}
which is zero. Hence, it is possible to define a 5D vacuum in the Weitzenb\"ock sense $^{(W)}R^a_{bcd} = 0$. Is easy to see that the space-time associated to the metric of the
equation (\ref{5ds22}) is not Ricci-flat in a Riemannian. WIMT is an interesting tool in the cases that we cannot apply IMT in a direct way. This is when $^{(5D\,R)}R^A_{BCD}\neq 0$. In this case it is possible to transform our problem to a Weitzenb\"{o}ck geometry with the equation (\ref{conerel}), obtaining that $^{(5D\,We)}R^A_{BCD}=0$,
in order to induce the effective tetra-manifold which is the representation of a space-time $4DM$ and the tensor objects linked to the physical elements in the manifold.

\subsection{Magnetic Monopoles in a Weitzenb\"{o}ck geometry}\label{ap3}

We consider the action ${\cal S}$, on a 5D non-vacuum Riemann manifold
\begin{equation}
{\cal S} = \int\,d^5x \,\sqrt{|g|} \,\left[ \frac{^{(5)} {\cal R}}{16\pi \,G} -\frac{1}{4} F_{ab} F^{ab} - J_b A^b\right],
\end{equation}
where $^{(5)} {\cal R}$ is the scalar curvature, $F^{ab} = \nabla^a A^b - \nabla^b A^a$ is the gravitoelectromagnetic tensor, and $J_b$ are the five components
of the gravitoelectromagnetic currents. The Maxwell equations, in the language of differential forms must be written as
\begin{eqnarray}\label{max}
*d(F)=\,^{(m)}J,\nonumber \\
*d(*F)=\,^{(e)}J.
\end{eqnarray}
Here, $F$ is the analog to the Faraday $2$-form, $d$ is the exterior covariant derivative and $*$ the adjunction or duality operation in dimension $k$.
The $p$-form is a tensor object $W$, of order $p$ and cotangent, which is anti-symmetric and is described by
\begin{eqnarray}\nonumber
W=\frac{1}{p!}\,w_{i_1\,...\,i_p}\,\underrightarrow{e}^{i_1}\wedge...\wedge\underrightarrow{e}^{i_p},
\end{eqnarray}
where the wedge product is the anti-symmetrization of the tensor product. The exterior covariant derivative associated to a covariant derivative "$;$", is
\begin{eqnarray}\nonumber
d(W)=\frac{1}{p!}\,w_{i_1\,...\,i_p\,;k}\,\underrightarrow{e}^k \wedge \underrightarrow{e}^{i_1}\wedge...\wedge\underrightarrow{e}^{i_p},
\end{eqnarray}
and the adjunction in a manifold of dimension $k$, is defined by
\begin{eqnarray}\nonumber *W=\frac{\sqrt{|g|}}{(k-p)!\,p!}\varepsilon_{j_1\,...\,j_pi_{p+1}\,...\,i_n}w^{j_1\,...\,j_p}\,&\underbrace{\underrightarrow{e}^{i_{p+1}}
\wedge...\wedge\underrightarrow{e}^{i_k}}&.\nonumber \\
&k-p&
\end{eqnarray}
The adjunction operation takes a $p$-form and produces a $(k-p)$-form. The Faraday $2$-form is defined from the exterior covariant derivative of a $1$-form which in its tangent shape is the a penta-vector $A=(\varphi,\overrightarrow{A})$
\begin{eqnarray}\nonumber
F=d(A).
\end{eqnarray}
In the case that the connection is symmetric, then $d(F)=d(d(A))=0$, which implies the absence of magnetic monopoles in the theory.
The source term $^{(m)}J$, is a cotangent vector (or co-vector) of the magnetic current $^{(e)}J$, which is the electric one.
In both cases $^{(*)}J_0=\rho_{*}$, the first component of the co-vector is the charge density.

All topics exposed in the present section must be directly interpreted in dimension $k=n+1$, and after making the foliation that leads us to the effective currents in
dimension $k-1=n$. In the present work we aboard the special case in which $n=4$ and $k=5$. From the equation (\ref{max}), we could see that the electric part of the equation $*d(*F)=\,^{(e)}J$, in $5D$, implies that the $5D$-dual of $F$ (remembering that $F$ is a $2$-form), is a $3$-form. The exterior covariant derivative of the obtained
$3$-form is a $4$-form and the adjoint or $5D$-dual of such $4$-form is a $1$-form, so that we obtain that $\underrightarrow{^{(5D\,e)}J}$ is a co-vector field. The foliation
conduces us to the effective co-vector field. On the other hand, for the magnetic part $*d(F)=\,^{(m)}J$, the exterior covariant derivative of a $2$-form is a $3$-form and his $5D$-dual is a $2$-form. In this case $\underrightarrow{\underrightarrow{^{(5D\,m)}J}}$ is a $2$-cotangent tensor field. Anyway, is not difficult to prove that such object contains
the same information that a co-vector field defined as
\begin{eqnarray}\label{cormag}
\underrightarrow{^{(5D\,m)}J}=\underrightarrow{\underrightarrow{^{(5D\,m)}J}}(\check{n},\,),
\end{eqnarray}
where $\check{n}$ is a vector field, tangent to 5DM, but normal to 4DM in each point. The equation (\ref{cormag}) grants that we obtain the right equations for the effective fields induced in $4DM$, after making the foliation.

\subsection{The 5D metric: Inner and Outer zones}

We consider a five dimensional manifold called 5DM with a
metric characterized by $g=g_{ab}\,dx^a \otimes dx^b$, which is
described by a coordinate basis of the cotangent CT5DM vector space. The basis is symbolized by $\{dx^a\}$,
\begin{footnote}{We consider that the indexes $a=0,1,2,3,4$ run over the 5 coordinates with index $0$ corresponding to the time. The coordinate basis for the tangent space 5DTM is $\{\overrightarrow{e}_a\}=\{\overrightarrow{\frac{\partial\,\,\,}{\partial\,x^{a}}}\}=\{\overrightarrow{\frac{\partial\,\,\,}{\partial\,t}},\overrightarrow{\frac{\partial\,\,\,}{\partial\,r}},
\overrightarrow{\frac{\partial\,\,\,}{\partial\,\theta}},\overrightarrow{\frac{\partial\,\,\,}{\partial\,\varphi}},\overrightarrow{\frac{\partial\,\,\,}{\partial\,\psi}}\}$ expressed in extended spherical coordinates for 5DM. \\We shall use small primed indexes $a'=0,1,2,3,4$ to denote another coordinate basis of 5DTM and big indices $A=0,1,2,3,4$ to make reference to an ortho-normalized basis of 5DTM, which in general must be non-coordinate.}
\end{footnote}
with the length element $^{(5D)}dS^2=g_{ab}\,dx^a\,dx^b$, according to
\begin{eqnarray}\label{5ds2}
^{(5D)}dS^2=
f(r)\,dt^2-f(r)^{-1}\,dr^2-r^2\,d\theta^2-r^2\,sin(\theta)^2\,d\varphi^2-d\psi^2,
\end{eqnarray}
where $r\geq 0$, and $f(r)$ given by
\begin{eqnarray}\label{f(r)}
f(r)=1-\frac{2\,m}{r}.
\end{eqnarray}
The critical value $r_{sch}=2m$, implies that $g_{00}\rightarrow 0$ and
$g_{11}\rightarrow\infty$, so that the manifold is separated in two different zones. We call "outer zone" for $r>2m$, and "inner zone" for $r<2m$.

\section{Inner Geometry}

In this case we consider the metric of (\ref{5ds2}) over the outer zone of the manifold. If we extend the definition in
(\ref{trafoext}), we obtain that $u^2:=r-2m$ and then the
coordinate $u$ assume complex values. To avoid such problem we use a new transformation defined according to
\begin{eqnarray}\label{trafoint}
u^2:=2m-r,
\end{eqnarray}
with $0\leq r < 2m$. Using the transformation (\ref{trafoint}) in the equation (\ref{5ds2}), we obtain
\begin{eqnarray}\label{5ds22}
^{(5D)}dS'^2&=&
\frac{u^2}{u^2-2m}\,dt^2-4(u^2-2m)\,du^2\nonumber \\
&- &\,(2m-u^2)^2\,(d\theta^2+\,sin(\theta)^2\,d\varphi^2)-d\psi^2,
\end{eqnarray}
which is the metric that describes the inner zone of $5DM$\begin{footnote}{In analogy with our development for the transformation (\ref{trafoext}), the new transformation (\ref{trafoint})
acts over the basis of $5D\,CTM$, in the inner zone. We named such basis $dx^a= \{dt,\, dr,\,d\theta,\,d\varphi,\,d\psi\}$ and $dx^{\bar{a}}= \{dt,\, du,\,d\theta,\,d\varphi,\,d\psi\}$, which must be related with the help of the vierbeins: $dx^a=e^a_{\bar{a}}\,dx^{\tilde{a}}$, where \begin{eqnarray}\label{vierint}
[e^a_{\bar{a}}]=\left(%
\begin{array}{ccccc}
  1 & 0 & 0 & 0 & 0 \\
  0 & -2u & 0 & 0 & 0 \\
  0 & 0 & 1 & 0 & 0 \\
  0 & 0 & 0 & 1 & 0 \\
  0 & 0 & 0 & 0 & 1 \\
\end{array}%
\right).\end{eqnarray}
Must be remarked that this expression is only valid for the inner
zone.}\end{footnote}.
Now, we shall study the effective metric obtained for $4DM_{int}$, by making some constant foliations.

\subsection{Foliation $\theta=\theta_0$}\label{titacero}

Our interest is centered in the foliations of the
$\theta=\theta_0$ type which grants that $4DM_T$ is at least
multiple connected (must have holes), but representing a true nexus
between different outer zones which are effective BH
$4DM_{ext\,\,\psi_i}$ in the outer geometry.  If we consider the foliation $\theta=\theta_0=\frac{\pi}{2}$ on the metric (\ref{5ds22})
\begin{eqnarray}\label{4ds23}^{(4D)}dS'^2&=&\,^{(5D)}dS'^2|_{\,\theta=\frac{\pi}{2}}=\\\nonumber
&=&
\frac{u^2}{u^2-2m}\,dt^2-4(u^2-2m)\,du^2-\,(2m-u^2)^2\,d\varphi^2-d\psi^2.\end{eqnarray}
The effective inner metric will describe an effective manifold $4DM_{int}$ which is Lorentzian and invertible.
Furthermore, this manifold could be a nexus between different Einstein-Rosen
bridges or Schwarzschild BH, obtained from different constant
foliations over $\psi$ for the \textbf{outer zone}\footnote{An interesting case by which cannot is possible to
obtain a wormhole is that in which the foliation is $\psi=\psi_0$. This case is studied in the appendix \ref{app2}.}.

The geodesic curves for the inner zone are described by
\begin{eqnarray}\label{geo}
\frac{\partial^2 \varphi}{\partial s^2}&-&4\frac{u}{2m-u^2}\frac{\partial u}{\partial s}\frac{\partial \varphi}{\partial s}=0,\\
\frac{\partial^2 t}{\partial s^2}&+&4\frac{m}{(2m-u^2) u}\frac{\partial u}{\partial s}\frac{\partial t}{\partial s}=0,\\\frac{\partial^2 u}{\partial s^2}+\frac{1}{2}\frac{m u}{(2m-u^2)^3 }\left(\frac{\partial t}{\partial s}\right)^2&-&\frac{u}{2m-u^2}\left(\frac{\partial t}{\partial u}\right)^2-\frac{u}{2}\left(\frac{\partial \varphi}{\partial s}\right)^2=0,\\
\frac{\partial^2 \psi }{\partial
s^2}&=&0,
\end{eqnarray}
in which $s$ is the length of the effective $4DM$ line element. The simplest solution is $t=t_0,\,\,u=u_0$, with $\varphi=\varphi_0,\,\,\psi=\psi_0+ks$,
which, when $k=0$ is reduced to the trivial case in which any coordinate remains constant. The most interesting case is that in which, the magnetic charge rotates around the compact coordinate $\varphi$ advancing along the extra dimension $\psi$. This helicoidal geodesics along the extra coordinate can be identified as the one of a magnetic charged particle moving into a traversable magnetic field, with trajectory:
\begin{eqnarray}\label{geosol}
\psi (s)&=&\psi_0+Ks,\\\varphi (s) &=& \varphi_0 e^{iRs},\\ u(s)&=&\pm\sqrt{2m\left(1- N e^{-\frac{1}{2}R^2s}\right)},\\t(s)&=&t_0 \left(1+e^{-\frac{1}{2}K^2 s}\right).
\end{eqnarray}
Here, the constants $K$ and $R$, are constants of integration, which are supposed to be real. The constant $N$ is in the range $0\leq N<1$, because this condition is necessary
and sufficient to obtain that $-\sqrt{2m}<u<+\sqrt{2m}$, the definition range of $u$. We must notice that $u\,\rightarrow\, 2m$ when ${s\,\rightarrow\,\infty}$,
which is the value of the coordinate associated to $r=0$. Then, by making the assumption that $|K| \gg R^2$, we must view the geodesics as helical trajectories with axis along $\psi$, with a value of $u$ varying very slowly, and making revolutions in compact coordinate $\varphi$. Under the last assumptions the geodesic trajectory of a charged particle is approximately helical and very similar to the "string" of a Dirac monopole, but inside the wormhole. Curiously, the affine parameter $s$ is a strictly decreasing function of the time and vice versa; the time $t_0$ is "reached" for $s\,\rightarrow \infty$, where $\frac{t_0}{2m}$ is an integration constant. An interesting question must arise: Could be the effective exterior electric charge the effect of the inner monopole traveling along an helical trajectory in the wormhole? Of course, this phenomena cannot be as a consequence of causality, because the inner and the exterior regions are causally disconnected, but can be understood from the point of view of a topological induction, because both regions (the inner and the outer), are obtained as different foliations of an unique 5D spacetime. Finally, as we shall see later [see, eq. (\ref{doduo})], this topological choice, supported by the experimental evidence, in which both charges cannot coexist in causally connected regions, makes possible the quantization of the gravito-electric and gravito-magnetic charges.

\subsection{Gravito-Magnetic Currents for the inner observer.}

We consider the transformation that relates the
coordinate basis of 5D CTM bridge $\{dx^{\bar{a}}\}$ with an
ortho-normalized, (non-coordinate) basis denoted by $\{e^{A}\}$: $e^{A}=\bar{e}^A_{\bar{a}}\,dx^{\bar{a}}$
\begin{eqnarray}\label{viernorm}[\bar{e}^A_{\bar{a}}]=\left(%
\begin{array}{ccccc}
  -\frac{u}{\sqrt{u^2-2m}} & 0 & 0 & 0 & 0 \\
  0 & 2\sqrt{2m-u^2} & 0 & 0 & 0 \\
  0 & 0 & 2m-u^2 & 0 & 0 \\
  0 & 0 & 0 & (2m-u^2)\sin\theta & 0 \\
  0 & 0 & 0 & 0 & 1 \\
\end{array}%
\right).\end{eqnarray}
The Weitzenb\"{o}ck connections can be obtained using the equations (\ref{conewei}) and (\ref{conewei1}). By checking
the zero curvature tensor and calculating the torsion, we obtain that the anti-symmetric components $^{(5D\,We)}T^A_{BC}$, are
\begin{eqnarray}\label{thor}
&^{(5D\,We)}T^{0}_{01}=&-\frac{1}{2}
\left(\frac{1}{u\sqrt{2m-u^2}}+\frac{u}{(2m-u^2)^{3/2}}\right),\nonumber \\
&^{(5D\,We)}T^{2}_{21}=&\frac{u}{(2m-u^2)^{3/2}},\nonumber \\
&^{(5D\,We)}T^{3}_{31}=&\frac{u}{(2m-u^2)^{3/2}},\nonumber \\
&^{(5D\,We)}T^{3}_{32}=&-\frac{1}{2m-u^2}\frac{\cos\theta}{\sin\theta}.
\end{eqnarray}
This is valid for $\sin\theta\neq0$. Otherwise, the torsion tensor coefficients are null. Furthermore, by using the (\ref{tor1}) it is easy to obtain the contortion tensor (\ref{contor}) and all the components $^{(5D\,We)}T^{\bar{a}}_{\bar{b}\bar{c}}$. According to (\ref{max}), the form of the magnetic $5D$ currents in the Weitzenb\"{o}ck geometry is
\begin{equation}\label{2formacorrmag}
[*\left(d(F)\right)]^{AB}=\frac{1}{3}\,\varepsilon^{ABCDE}\mathcal{J}_{DEC},\end{equation}
with
\begin{eqnarray}\nonumber \mathcal{J}_{DEC}&=&\,^{(5D\,We)}T^{K}_{DC}E_K(A_E)+\,^{(5D\,We)}T^{K}_{CE}E_K(A_D)+\,^{(5D\,We)}T^{K}_{ED}E_K(A_C)\nonumber \\ &\,&+E_C(\,^{(5D\,We)}T^{N}_{ED}A_N)+E_E(\,^{(5D\,We)}T^{N}_{DC}A_N)+E_D(\,^{(5D\,We)}T^{N}_{CE}A_N).
\end{eqnarray}
Here, we have applied WIMT with the inner and constant foliation characterized by $\theta=\theta_0=\frac{\pi}{2}$, which is in agreement with the scenario of (\ref{titacero}). Hence, the effective magnetic current is
\begin{eqnarray}\label{1formacorrmag}
^{(4D\,m)}\overrightarrow{\mathcal{J}}=\frac{1}{3!}\left.\varepsilon^{a2cde}
\,\overrightarrow{e}_a\,\mathcal{J}_{2cde}\right|_{\theta=\frac{\pi}{2}},
\end{eqnarray}
with components: $^{(4D\,m)}{\mathcal{J}}^I=\frac{1}{3!}\left.
\bar{e}^I_a\,\varepsilon^{a2cde}\mathcal{J}_{2cde}\right|_{\theta=\frac{\pi}{2}}$.
Here, we have used (\ref{cormag}), taking into account that the
vector is normal to the $4DM_{int}$ points in the
$\theta$-direction. Once applied the foliation, we obtain that
\begin{eqnarray}\label{denscarmagint}
\varrho_m=\frac{2u^2}{(2m-u^2)^{3/2}}\left(\frac{\partial\,A_{4}}{\partial\varphi}+\frac{\partial\,A_3}{\partial\psi}\right),
\end{eqnarray}
\begin{eqnarray}\label{jfi}\mathcal{J}_{\varphi}=\frac{u}{(2m-u^2)^{3/2}} \frac{\partial A_{0}}{\partial \varphi} +\frac{1}{2} \left(\frac{1}{u^2}+\frac{1}{2m-u^2}\right)
\frac{\partial A_{2}}{\partial t}.
\end{eqnarray}
The expression (\ref{denscarmagint}) is the gravito-magnetic
charge density,  and (\ref{jfi}) represents the associated current
in the $\varphi$-direction. The current is similar to whole in a
coil along the extra direction $\psi$,  with $\mathcal{J}_{u}=0\,$
and $\,\mathcal{J}_{\varphi} \gg \mathcal{J}_{\psi} \cong 0$. This
is compatible with helical geodesic present in previous section.

\section{Outer geometry: Effective 4D Schwarzschild BH}

In the outer zone we are dealing with the usual coordinate transformation, which characterizes the Einstein-Rosen bridge. Such transformation is given by
\begin{eqnarray}\label{trafoext}
v^2:=r-2m.
\end{eqnarray}
On the outer zone $r-2m$ is a positive number and $v=\pm \sqrt{r-2m}$ allows us to obtain two different and possible values
of $v$ for any $r>2m$, so that $-\infty < v < \infty$. The value $v=0$ is obtained in $r=r_{sch}=2m$ by removing the inner
zone. We obtained a fifth-dimensional extension of the Einstein-Rosen bridge in the sense that, if we apply a constant
foliation $\psi=\psi_0$ on the metric (\ref{5ds2}), we recover a fourth-dimensional manifold with the usual metric for a wormhole.

After making the transformation (\ref{trafoext}) for the outer zone, we obtain that the metric (\ref{5ds2}) becomes\begin{footnote}{Transformation (\ref{trafoext}) acts over the basis of $5DCTM$ in the outer zone. We call respectively, original and transformed basis: $dx^a= \{dt,\, dr,\,d\theta,\,d\varphi,\,d\psi\}$ and $dx^{\tilde{a}}= \{dt,\, dv,\,d\theta,\,d\varphi,\,d\psi\}$, where the vierbein $e^a_{\tilde{a}}$, help us to transform a coordinate in the other: $dx^a=e^a_{\tilde{a}}\,dx^{\tilde{a}}$, with \begin{eqnarray}\label{vierext}
[e^a_{\tilde{a}}]=\left(%
\begin{array}{ccccc}
  1 & 0 & 0 & 0 & 0 \\
  0 & 2v & 0 & 0 & 0 \\
  0 & 0 & 1 & 0 & 0 \\
  0 & 0 & 0 & 1 & 0 \\
  0 & 0 & 0 & 0 & 1 \\
\end{array}%
\right).
\end{eqnarray}
Notice that this expression and treatment is only for outer zone.}
\end{footnote}
\begin{eqnarray}\label{5ds21}
^{(5D)}dS'^2&=& \frac{v^2}{v^2+2m}\,dt^2-4(v^2+2m)\,dv^2 \nonumber \\
&- & \,(2m+v^2)^2\,(d\theta^2+\,sin(\theta)^2\,d\varphi^2)-d\psi^2.
\end{eqnarray}
After making the foliation $\psi=\psi_0$ in the metric (\ref{5ds21}), we obtain
\begin{eqnarray}\label{4ds21}
^{(4D)}dS'^2&=&\,^{(5D)}dS'^2|_{\psi=\psi_0} \nonumber \\
&=& \frac{v^2}{v^2+2m}\,dt^2-4(v^2+2m)\,dv^2-\,(2m+v^2)^2\,(d\theta^2+\,sin(\theta)^2\,d\varphi^2),
\end{eqnarray}
which is the usual Einstein-Rosen metric that describes a space-time bridge. Now, we can define the area of the spherical shell related to the
effective wormhole neck, described by (\ref{4ds21})
\begin{eqnarray}\label{aext}
A_{ext}(v)=4\,\pi \,r(v)^2,
\end{eqnarray}
where $r(v)=v^2+2m$. Hence, taking into account (\ref{trafoext}), the area takes minimum value at $v=0$
\begin{eqnarray}\label{aminext}
A_{min}=16\,\pi\,m^2.
\end{eqnarray}
The maximum area is not bounded and is reached over the asymptotic planes. On the other hand, $v \rightarrow
\pm\infty$, and it is easy to see that $^{(4D)}dS'^2 \rightarrow \,^{(4D)}dS'^2|_{Minkowsky}$. Since $v$ is bi-valuated for each $r$, there are two plane spaces located for the extreme
values of $v$. They are separated by a bridge with variable neck size, according to (\ref{aext}). We remark that the transformation
(\ref{trafoext}) over the exterior space-time causes that, in the
new coordinate, the inner zone is completely removed and the
entire boundary is reduced to a single point with $v=0$.

\subsection{Gravito-Magnetic Currents for the effective outer zone.}

Gravito-magnetic $4D$-effective currents for the outer metric must
be obtained from the paper\cite{nuestro} [see the equation (46), therein], and $^{(LC)(m)}J_0\mid_{\psi_0}=\rho_m$ takes form
\begin{eqnarray}\label{current0xxx}
\rho_m= \left\{-\frac{\sqrt{f(r)}}{r}\frac{\partial
A_3}{\partial\theta}+\frac{\sqrt{f(r)}}{r
\sin(\theta)}\frac{\partial
A_2}{\partial\varphi}-\frac{\cos(\theta)}{r^2
\sin^2(\theta)}\frac{\partial A_1}{\partial\varphi}\right\},
\end{eqnarray}
by normalizing a constant scale factor. In order to obtain the currents for the present case, we must set $f(r)=1-\frac{2M}{r}$
and $r=v^2+2m$ and transform to the bridge base $\{dx^{\tilde{A}}\}$ with the vierbeins. For the
exterior Wu-Yang potentials
\begin{footnote}{In a previous work we have checked the consistency of our equation of gravitomagnetic current for a 4D BH with the 4D Wu-Yang potentials. These ones describe a localized magnetic monopole with charge $q_m$ \begin{eqnarray}\nonumber
\overrightarrow{A}^{(N)}&=&q_m \frac{(1-\cos(\theta))}{r
\sin(\theta)}\overrightarrow{e}_\varphi=\frac{q_m}{r}A^{(N)}_3\overrightarrow{e}_{\varphi},\\
\nonumber \overrightarrow{A}^{(S)}&=& -q_m
\frac{\left(1+\cos(\theta)\right)}{r
\sin(\theta)}\overrightarrow{e}_\varphi=\frac{q_m}{r}A^{(S)}_3\overrightarrow{e}_{\varphi}.
\end{eqnarray}
Here, labels $(N)$ and $(S)$ indicate the North or South hemisphere, on which are valid the potentials.}
\end{footnote}[the readers could see \cite{wu}], we obtain that an exterior observer sees exterior fields
compatible with an inner gravito-magnetic density
\begin{eqnarray}\label{current0}
\rho_m=-\frac{2v^2}{(v^2+2m)^{3/2}}\frac{\partial
A_3}{\partial\theta}.
\end{eqnarray}
This expression (\ref{current0}) grants that, in the outer zone
the observer must perceives an effective gravito-magnetic
monopole, which is in general not zero. In association with
Wu-Yang potentials we must see that such gravito-magnetic charge
ensures the accomplishment of the Dirac quantization condition
$q_e\,q_m=\frac{n}{2}$, with $n \in \mathbb{Z}$.\\Now we are going
to follow the geometric product quantization presented in art.
\cite{dyonic} by doing $^{(ge)} \underrightarrow{J} \,
^{(gm)}\underrightarrow{J}=\, ^{(ge)} \underrightarrow{J} \, \cdot
\, ^{(gm)}\underrightarrow{J}+\, ^{(ge)} \underrightarrow{J} \,
\wedge \, ^{(gm)}\underrightarrow{J} $ which is the geometrical
product of the gravito-electric $1$-form and the gravito-magnetic
$1$-form of current resulting in the sum of an scalar and a
$2$-form. With $\left( ^{(ge)} \underrightarrow{J} \,
^{(gm)}\underrightarrow{J} \right)^2=n^2$ in which the square is
taken in the multi-tensor sense\begin{footnote}{A multi-tensor
object must be expressed as the sum of a scalar function and
tensors of different nature:
\begin{eqnarray}\label{t}
T=A+B_n \underrightarrow{e}^n+C_{nm} \underrightarrow{e}^n \otimes
\underrightarrow{e}^m + D_{nmp}\underrightarrow{e}^n \otimes
\underrightarrow{e}^m \otimes \underrightarrow{e}^p+...
\end{eqnarray}
where $\underrightarrow{e}^n$ is running over a basis of the
co-tangent space, $\otimes$ is the tensor product of such elements
and $A, B_n, C_{nm}, D_{nmp}$ are arbitrary scalar functions in
$\mathfrak{F}(M)$. We must obtain a scalar defined by
\begin{eqnarray}\label{t2}
T^2&=&A^2+ B_n B_{n'}
\overrightarrow{\overrightarrow{g}}(\underrightarrow{e}^n,\underrightarrow{e}^{n'})
+ C_{nm} C_{n'm'}
\overrightarrow{\overrightarrow{g}}(\underrightarrow{e}^n,\underrightarrow{e}^{n'})
\overrightarrow{\overrightarrow{g}}(\underrightarrow{e}^m,\underrightarrow{e}^{m'})
\\ \nonumber &\,& + D_{nmp} D_{n'm'p'}
\overrightarrow{\overrightarrow{g}}(\underrightarrow{e}^n,\underrightarrow{e}^{n'})
\overrightarrow{\overrightarrow{g}}(\underrightarrow{e}^m,\underrightarrow{e}^{m'})
\overrightarrow{\overrightarrow{g}}(\underrightarrow{e}^p,\underrightarrow{e}^{p'})
+ ...\\\nonumber &=& A^2 + B_n B_{n'} g^{nn'}+ C_{nm} C_{n'm'}
g^{nn'} g^{mm'} + D_{nmp} D_{n'm'p'} g^{nn'} g^{mm'} g^{pp'}+... ,
\end{eqnarray}
which is a generalized expression of a inner product for
multi-tensorial objects of any kind. We must notice that such
product coincides with the usual inner product for a pure vector
or co-vector.}
\end{footnote} and we obtain that
\begin{eqnarray}\label{cuant}\rho_e\,\rho_m=\frac{\sqrt{n^2-m^2}}{2},\end{eqnarray}
with $n>m$, $n$ and $m$ could be zero or integer positive numbers.
Is easy to see that $m$ must be zero in eq. (\ref{cuant}) in order
to agree with the Dirac quantization condition, this assertion is
equivalent to say that the gravito-magnetic and the
gravito-electric currents must be orthogonal. Such scenario is
including the trivial case in which $\rho_m=0$ and the most
general one in which $\rho_m\neq 0$ but the monopoles distribution
is static. \\In similar manner we must obtain an expression analog
to eq. (\ref{cuant}) for the inner zone.

In order to clarify the dual nature of the formalism and relate both, the inner and outer zones, we shall
suppose that the inner charge is completely magnetic. If we work using an ortho-normalized basis, with the Lorentz gauge in mind, we
obtain
\begin{eqnarray}\label{doduo}
\rho_{e}=\,\alpha\,
\varepsilon^{04CDE}\mathcal{J}_{DEC}.
\end{eqnarray}
Now, if we take into account the limit on the frontier between both zones, we obtain
\begin{eqnarray}\label{doduo}
\rho_{e}=\,\alpha\,\varrho_m,
\end{eqnarray}
where $\rho_e$ is the electrical density of charge in the outer zone and
$\varrho_m$ is the magnetic density of charge inside the wormhole. Furthermore, $\alpha$ is a factor of proportionality,
dependent of the coordinates.  Additionally, we have supposed that the electrical density of charge, $\varrho_e=0$, is zero  in
the inner zone, and then $\rho_m=0$ in the outer zone. This is a different kind of duality of the extended Einstein-Maxwell
equations, that relates electric and magnetic charges on causally disconnected space regions.

\section{Final Comments}

We have studied traversable wormholes over an effective
Schwarzschild space-time using a foliation $\theta=\pi/2$ on an
extended 5D non-vacuum space-time. In this space-time, the
electrodynamics can be extended to a gravito-electrodynamics
theory, described by a $2$-form $F_{ab}$ tensor and a $3$-form
${\cal F}_{abc}$ dual tensor. In sections II and III we study the
gravito-magnetic currents in the frame of WIMT and the induced
geodesics which must take the form of an helical trajectory in the
inner zone for a charged particle, then we make a question "Could
be the effective exterior electric charge the effect of the inner
monopole traveling along an helical trajectory in the wormhole?".
At the end of section IV A we present the possibility that the
gravito-electric and gravito-magnetic charges are causally
disconnected and positively answering to the question formulated
at the end of section III A, in the case where gravito-electric and
gravito-magnetic currents are orthogonal. The associated dynamics
of the gravito-electromagnetic potentials $A_b$ make possible to
describe gravito-magnetic currents in presence of a stable
gravito-magnetic monopole in the interior of the wormhole, but the
relativistic observer which is placed outside the BH must be not
sensitive to the gravito-magnetic charge inside the BH, perceiving
only the effective electric charge. This effect could be possible because both kinds
of charges (the gravito-electric and gravito-magnetic ones) are topologically induced from a
unique 5D spacetime. In particular, the
quantization condition arises as an effective manifestation of the
current expression between the components of the penta-velocity of
the observer. At the beginning of section III A we have obtained
an induced gravito-magnetic monopole in the framework of WIMT
which must be in general non zero and compatible with the Wu-Yang
fields, with the same magnitude of charge and quantization of
charge included. This effective exterior monopole is treated under
geometric quantization which is reduced to the Dirac quantization
condition in the static case.

\section*{Acknowledgements}

\noindent J. M. Romero and M. Bellini acknowledge CONICET
(Argentina) and UNMdP for financial support.

\appendix

\section{Weitzenb\"{o}ck Geometry}\label{ap1}

The Weitzenb\"{o}ck geometry must be constructed from the vierbeins,
which are the coefficients that express the relation between $\{\overrightarrow{E}_A\}$ and $\{\overrightarrow{e}_a\}$. These are two different basis of $TM$.
\begin{eqnarray}\label{vier}
\overrightarrow{E}_A=e_A^a\,\overrightarrow{e}_a,\,\,\,\,\,\,\overrightarrow{e}_a=\bar{e}_a^A\,\overrightarrow{E}_A.
\end{eqnarray}
The vierbeins comply with the property
\begin{eqnarray}\label{vier2 }
e_A^a\,\bar{e}^A_b=\delta^a_b,\,\,\,\,\,e_B^b\,\bar{e}^A_b=\delta^A_B.
\end{eqnarray}
The transformation of any arbitrary tensor $T$ in ${\mathcal{T}^p}_m(M)$, from one basis to another is
\begin{eqnarray}\label{trafo}
{T^{a_1\,...\,a_p}}_{b_1\,...\,b_m}=e^{a_1}_{A_1}\,...\,e^{a_p}_{A_p}\,\bar{e}^{B_1}_{b_1}\,...\,\bar{e}^{B_m}_{b_m}{T^{A_1\,...\,A_p}}_{B_1\,...\,B_m}.
\end{eqnarray}
Furthermore, the Weitzenb\"{o}ck connections are defined by
\begin{eqnarray}\label{conewei}
^{(We)}\Gamma^a_{bc}=e^a_N\overrightarrow{e}_c(\bar{e}^N_b).
\end{eqnarray}
In this way is defined a very special parallel transport condition founded in:
\begin{eqnarray}\label{vier2xxx }
{e_A^a}_{;a}=0.
\end{eqnarray}
Following the definition (\ref{conewei}), we obtain
\begin{eqnarray}\label{conewei1}
^{(We)}\Gamma^A_{BC}=0,
\end{eqnarray}
and then
\begin{eqnarray}\label{weR^A_BCD}
^{(We)}R^A_{BCD}=0.
\end{eqnarray}
In the usual Weitzenb\"{o}ck scenario the arrival basis $\{\overrightarrow{e}_a\}$ is a coordinate basis of the $TM$, with certain metric $g_{ab}$ of our interest. The departure basis
$\{\overrightarrow{E}_A\}$, must be non-coordinate but ortho-normalized. Then, the metric components in such basis are $\eta_{AB}$. In such context the torsion is
\begin{eqnarray}\label{tor1}
^{(We)}T^a_{bc}=e^a_A\,\bar{e}^B_b\,\bar{e}^C_c\,^{(We)}T^A_{BC}=e^a_A\,\bar{e}^B_b\,\bar{e}^C_c\,C^A_{BC},
\end{eqnarray}
where $C^A_{BC}$ are the structure coefficients of the basis $\{\overrightarrow{E}_A\}$, that comply with the algebra
\begin{eqnarray}\label{estruc}
[\overrightarrow{E}_B,\overrightarrow{E}_A]=C^C_{AB}\,\overrightarrow{E}_C.
\end{eqnarray}
The non-metricity of the connection is
\begin{eqnarray}\label{nomewe}
^{(We)}N_{abc}=\bar{e}^A_a\,\bar{e}^B_b\,\bar{e}^C_c\,^{(We)}N_{ABC}=\bar{e}^A_a\,\bar{e}^B_b\,\bar{e}^C_c\,\eta_{AB\,,C},
\end{eqnarray}
where we have used (\ref{tor1}), and in (\ref{nomewe}) we have employed the equation (\ref{conewei1}). It is usually used the ortho-normalized basis $\{\overrightarrow{E}_A\}$. Then,
due to the fact that $\eta_{AB}=-1,0,+1$ and $\eta_{AB\,,C}=0$, from the equation (\ref{nomewe}), we obtain
\begin{eqnarray}\label{nomewe2}
^{(We)}N_{ABC}=0.
\end{eqnarray}
All the earlier exposed items characterize the Weitzenb\"{o}ck geometry as a torsional geometry (\ref{tor1}), with (in usual cases zero) the non-metricity (\ref{nomewe2}), and a Weitzenb\"{o}ck-Riemann flat curvature tensor (\ref{weR^A_BCD}). The last assertion is fundamental for WIMT. The Weitzenb\"{o}ck connections and the Levi-Civita ones are related
by
\begin{eqnarray}\label{conerel}
^{(We)}\Gamma^A_{BC}=\,^{(R)}\Gamma^A_{BC}+\,^{(We)}K^A_{BC},
\end{eqnarray}
where $^{(We)}K^A_{BC}$ is the contortion tensor, which in the present case is enterally torsional due to the equations (\ref{tor1}) and
(\ref{nomewe2}). They take the form
\begin{eqnarray}\label{contor}
^{(We)}K^{A}_{BC}=-\frac{g^{AM}}{2}\{^{(We)}T^L_{CM}\,g_{BL}+\,^{(We)}T^L_{BM}\,g_{LC}-\,^{(We)}T^L_{CB}\,g_{ML}\}.
\end{eqnarray}

\section{Inner zone: foliation $\psi=\psi_0$}\label{app2}

We consider the metric (\ref{5ds22}) that describes the inner zone. If we make the foliation $\psi=\psi_0$, we
obtain
\begin{eqnarray}\label{4ds22}
^{(4D)}dS'^2&=&\,^{(5D)}dS'^2|_{\psi=\psi_0} \nonumber \\
&=& \frac{u^2}{u^2-2m}\,dt^2-4(u^2-2m)\,du^2-\,(2m-u^2)^2\,(d\theta^2+\,sin(\theta)^2\,d\varphi^2),
\end{eqnarray}
which represents the inner metric for the Einstein-Rosen wormhole. In present case $-\sqrt{2m}\leq u\leq+\sqrt{2m}$, and the new
coordinate takes is $u\rightarrow 0$
when $r \rightarrow r_{sch}$. On the other hand for $r=0$, $u$ is bi-valuated: $u(r=0)=\pm \sqrt{2m}$. The spherical area
associated to the neck of the wormhole is defined in analogy to the equation (\ref{aext})
\begin{eqnarray}\label{aint}
A_{int}(u)=4\,\pi \,r(u)^2,
\end{eqnarray}
but taking into account (\ref{trafoint}). Therefore, we obtain $r(u)=2m-u^2$. For the minimum value of
$u=-\sqrt{2m}$ the area of the neck also reaches a minimum at
\begin{eqnarray}\label{aintmin0}
A_{int}(u=-\sqrt{2m})=0.
\end{eqnarray}
It grows until the maximum value $u=0$, given by
\begin{eqnarray}\label{aintmax}
A_{int}(u=0)=16\,\pi\,m^2,
\end{eqnarray}
which is the same value than whole of (\ref{aminext}) for the outer minimum. Hence, it decrease until a minimum for the maximum value of $u=+\sqrt{2m}$:
\begin{eqnarray}\label{aintmin1}
A_{int}(u=+\sqrt{2m})=0.
\end{eqnarray}
The inner geometry is the same as in a capsule localized in $\psi=\psi_0$.

\end{document}